\documentclass[aps,pre,twocolumn,groupedaddress,showkeys,floatfix]{revtex4-1}

\usepackage{multirow}
\usepackage{color}
\usepackage{amsmath}
\usepackage[colorinlistoftodos,textsize=tiny]{todonotes}
\usepackage{epstopdf}
\usepackage{multirow}
\usepackage{subcaption}
\usetikzlibrary{patterns}
\usepackage{pgfplots}
\usepgfplotslibrary{external} 
\graphicspath{ {./figures/} }

\renewcommand{\vec}[1]{\mathbf{#1}}
\newcommand{\x}[0]{\vec{x}}
\renewcommand{\c}[0]{\vec{c}}

\begin{document}


\title{Curvature estimation from a volume of fluid indicator function for the simulation of surface tension and wetting with a free surface lattice Boltzmann method}


\author{Simon Bogner}
\email[]{simon.bogner@fau.de}
\author{Ulrich R{\"u}de}
\email[]{ulrich.ruede@fau.de}
\homepage[]{http://www10.informatik.uni-erlangen.de}
\affiliation{Lehrstuhl f{\"u}r Systemsimulation, Universit{\"a}t Erlangen-N{\"u}rnberg, Cauerstra{\ss}e 11, 91054 Erlangen, Germany}

\author{Jens Harting}
\email[]{j.harting@fz-juelich.de}
\homepage[]{http://mtp.phys.tue.nl/}
\affiliation{Forschungszentrum J{\"u}lich, Helmholtz-Institut Erlangen-N{\"u}rnberg (IEK-11), F{\"u}rther Stra{\ss}e 248, 90429 N{\"u}rnberg, Germany}
\affiliation{Department of Applied Physics, Technische Universiteit Eindhoven, P.O. Box 513, 5600 MB Eindhoven, The Netherlands}


\date{\today}

\begin{abstract}
The free surface lattice Boltzmann method (FSLBM) is a combination of the
hydrodynamic lattice Boltzmann method (LBM) with a volume of fluid (VOF)
interface capturing technique for the simulation of incompressible free surface
flows. Capillary effects are modeled by extracting the curvature of the
interface from the VOF indicator function and imposing a pressure jump at the
free boundary. However, obtaining accurate curvature estimates from a VOF
description can introduce significant errors. This article reports numerical
results for three different surface tension models in standard test cases, and
compares the according errors in the velocity field (spurious currents).
Furthermore, the FSLBM is shown to be suited to simulate wetting effects at
solid boundaries. To this end, a new method is developed to represent wetting
boundary conditions in a least squares curvature reconstruction technique. The
main limitations of the current FSLBM are analyzed and are found to be caused by
its simplified advection scheme.  Possible improvements are suggested.



\end{abstract}


\maketitle

\section{\label{sec:intro}Introduction}
The free surface lattice Boltzmann method (FSLBM) \cite{KoernerEtAl}
is a numerical method for the simulation of free surface flows
combining a volume of fluid (VOF) approach
\cite{HirtNichols,ScardovelliZaleski,Tryggvason2011} for interface
tracking with the lattice Boltzmann method (LBM)
\cite{BenziEtAl,ChenDoolen1998,AidunClausen,Wolf-Gladrow,Succi} for
hydrodynamics. We use the same definition of free surface flow as
described in \cite{ScardovelliZaleski,Veldman2007} where it denotes a
single phase flow problem containing free boundaries instead of a two
phase flow problem. VOF methods follow the notion of a sharp interface
representation, i.e., assuming hydrodynamic equations for the bulk of
the flow and modeling the interface by boundary conditions or by a
jump of flow parameters in \emph{one-fluid approaches} for two-phase
flows. This is in contrast to currently popular lattice Boltzmann
multiphase approaches \cite{ChenDoolen1998,Nourgaliev2003,Liu2014}
(e.g., color gradient model \cite{Gunstensen91}, Shan-Chen model
\cite{Shan93}, free energy model \cite{Swift95}), that are based on a
\emph{diffusive} interface assumption. In the FSLBM, the LBM is used
only to approximate the incompressible Navier-Stokes equations for the
liquid phase. With sharp interface simulation techniques capillary
effects need to be modeled explicitly in addition to the interface
tracking. Altogether, there are three components that the total
accuracy of the method depends on: the hydrodynamic solver (LBM), the
interface tracking (advection of indicator function), and the surface
tension model. While the accuracy of the LBM is well-understood
\cite{Frisch87,Junk2005,Holdych2004}, only very few results have been
reported addressing the accuracy of the FSLBM's advection scheme or
its validity to simulate surface tension. Nevertheless, the approach
has found numerous applications in surface tension driven complex flow
scenarios \cite{XingEtAl2007,Anderl2014b,Attar2011,Svec2012,Ammer2014}
and even high Reynolds number flows
\cite{Janssen2012,JanssenOffshore2010}. Hence, in the present paper we
evaluate the accuracy of the FSLBM in surface tension driven flows,
and also discuss existing limitations due to the original advection
scheme. While conventional VOF implementations rely on \emph{geometric
  reconstruction} to approximate the advection equation of the
indicator function \cite{ScardovelliZaleski,Tryggvason2011}, the FSLBM
instead exploits the specific nature of the LBM.  However, the present
results in this work indicate that though this simplification reduces
the algorithmic complexity significantly, it also comes with a
comparably low accuracy.

The simulation of surface tension involves the extraction of curvature
information from the interface defined through the fill level function
\cite{Brackbill92,Lafaurie94,ScardovelliZaleski,ZaleskiEtAl2009}. The
VOF indicator function is non-smooth by definition, making the
estimation of the interface curvature a non-trivial task. Numerical
errors in the determination of interface stresses lead to the effect
of undesired \emph{spurious currents} that can invalidate or
destabilize the 
computed solutions. We remark, that, if the surface
stress is included as force term in the momentum equation as in
\cite{Brackbill92}, then the incompatible discretization of pressure and
indicator function gradients becomes an additional source for spurious
currents \cite{Francois2006}. The present model does not have this
problem, because surface tension is imposed as a pressure jump at the
boundary instead. Anyway, errors in the curvature estimation are
reported 
as problematic -- primarily because the grid convergence is
poor 
unless more sophisticated approaches are employed. Modern VOF codes
therefore rely on higher order curvature estimation techniques to
suppress this error \cite{ZaleskiEtAl2009,Gerlach2006}. However, these
techniques require the use of larger stencils. 
This may be undesirable or is even impossible if only next-neighbor
information is available, as e.g.~in some parallel computing
environments. 
Hence, in this paper we review and present three different techniques
to approximate the interface capillary tension from the fill level
function data of a free surface code that are using local $3 \times 3
\times 3$ neighborhoods only.

The first method is an adaption of the classic \emph{finite
  difference} (FD) model by Brackbill et al. \cite{Brackbill92} that
uses finite difference computations to obtain the interfacial
curvature. As reported previously for this method, the magnitude of
the spurious currents makes it impossible to simulate small capillary
number flows correctly. Notice 
that in \cite{Brackbill92}, the surface tension appears as a force
term in the momentum equation, and further 
that this force is smoothed out over several grid cells. Hence this
model is also called \emph{continuous surface force} model in the
literature.  In the FSLBM presented here, surface tension is included
in the boundary condition instead.

The second model locally reconstructs the interface described by the
VOF function as a continuous surface made up of triangles. From this
\emph{triangular reconstruction} (TR) the curvature information can be
extracted. This method has first been presented in \cite{Pohl2007},
and turns out to effectively reduce the error due to spurious
currents, because the predicted curvature values are much more
accurate.

The third method also involves a local reconstruction of the interface
geometry. Based on a least squares approach a parabolic approximation
to the interface is constructed in a local neighborhood of cells. From
this \emph{least squares reconstruction} (LSQR) one obtains curvature
estimates with a high accuracy. A similar approach has also been
described in \cite{Popinet2009} as a fallback solution for a higher
order height-function technique. For the current paper, we have
extended the approach to problems including adhesive boundary
conditions. It is the only model in this study achieving a second
order rate of convergence in the classic spherical bubble benchmark.

We also evaluate the boundary conditions needed to simulate the
wetting behavior of solid surfaces for the FD and LSQR model. The
original work by Brackbill et al. \cite{Brackbill92} discusses
adhesive boundary conditions, which we adopted to the FSLBM context
for this paper. Since curvature estimation based on finite differences
often introduces larger errors, smoothing and filtering techniques are
typically used and have been extended to adhesive boundaries in
\cite{Raeini2012} for the simulation of low Capillary number flows.
Motivated also by the limited accuracy of the finite difference
approximation of curvature by finite differences, \cite{Veldman2007}
and \cite{Afkhami2008} switched to the height function technique
\cite{Sussman2007,Cummins2005} for the simulation of surface wetting.
Some other approaches use level sets instead of the VOF method to
represent the interface \cite{Liu2005}, or directly use non-Eulerian
techniques (e.g. \cite{Tryggvason2001}). In this paper, we present the
results obtained from the FSLBM in several simulations of wetting
surfaces while studying alternative 
surface tension models. This allows a
comparison between the FD approach to surface tension and the newly
developed LSQR
method. 


\section{\label{sec:method}Numerical Method}
\renewcommand{\c}[0]{\vec{c}}
Throughout this paper we assume a three-dimensional D$3$Q$19$ -
lattice Boltzmann model \cite{Wolf-Gladrow,QianEtAl1992} with $Q=19$
lattice velocity vectors $\vec{c}_q$ with $q=0, .., Q-1$.  We denote
the LBM data (discrete \emph{particle distribution function} - PDF) by
$\vec{f} = (f_0, f_1, .., f_{Q-1})$. The LBM data is defined for every
node within the liquid subdomain $\Omega(t)$, and follows the
evolution equation
\begin{subequations}
  \begin{align}
    f_q(\x+\c_q, t+1) &= f'_q(\x, t) \label{eq:lbe1},\\
    \vec{f}'(\x, t) &= \vec{f}(\x,t) +
    \vec{C}(\vec{f}(\x,t)) \label{eq:lbe2}.
  \end{align}
\end{subequations}
where $\vec{C}$ is a \emph{collision operator}, and $\vec{f}'$ is the
\emph{post-collision} distribution function. For the present paper, we
have used the hydrodynamic \emph{two relaxation time} (TRT) collision
operator \cite{Ginzburg2007}. The TRT operator has two eigenvalues
$\lambda_e$, $\lambda_o$, controlling the relaxation of the even and
odd parts of the PDF, respectively. Similar to the popular LBGK model
\cite{QianEtAl1992} the relaxation rate $\tau = - 1/\lambda_e$
controls the kinematic viscosity $\nu = c_s^2(\tau-1/2)$, with
$c_s=1/\sqrt{3}$ for the present D$3$Q$19$ model. The second parameter
$\lambda_o$ is chosen to minimize the error at straight axis-aligned
walls. The macroscopic variables of pressure $P(\x,t) = c_s^2
\rho(\x,t)$ and velocity $\vec{u}(\x,t)$ are defined via moments of
the distribution function,
\begin{subequations}
  \begin{align}
    \rho &= \sum_{q=0}^{Q-1} f_q, \\
    \rho \vec{u} &= \sum_{q=0}^{Q-1} \c_q f_q,
  \end{align}
\end{subequations}
at the respective node $\x$.

The FSLBM has first been described in \cite{KoernerEtAl}. To track the
interface position in simulations, a VOF approach introduces a
\emph{fill level} function $\varphi$ following an advection equation
over time. The function $\varphi$ is defined for each finite volume
(or \emph{cell}) as the volume fraction filled with liquid. Only
the flow within the liquid subdomain $\Omega(t)$, consisting of all
cells $\x$ with positive fill level $\varphi(\x)>0$, is simulated by
means of the LBM. The remaining cells, referred to as \emph{gas}
cells, become temporarily inactive. Within the liquid subdomain
$\Omega(t)$, we distinguish between \emph{liquid} and \emph{interface}
cells. To count as interface, the cell must have both liquid and gas
cells in the direct neighborhood defined by the lattice model. Only
the interface cells are allowed to have a fill level between zero and
one. The interface cells are also used to compute the advection of the
free surface in terms of the fill level function, and to impose a free
surface boundary condition on the LBM. In comparison to other VOF
methods, where the advection of the fill level function is a non-trivial
problem that needs to be solved in addition to the hydrodynamics, the
FSLBM exploits the nature of the lattice Boltzmann equation,
Eq.~(\ref{eq:lbe1}) and Eq.~(\ref{eq:lbe2}), to update the fill levels
directly \cite{KoernerEtAl}, and sets
\begin{subequations}
  \begin{equation}
    \begin{split}
      \varphi(\x, t+1) = & \\ 
       \varphi(\x, t) +& \frac{\sum_{q=1}^{Q-1} k(\x, q) \cdot (f'_{\bar{q}}(\x+\c_q,t) - f'_q(\x,t))}{\rho(\x,t+1)},
      \label{eq:advection1}
    \end{split}
  \end{equation}
where the coefficient $k(\x,q)$ is
\begin{equation}
  k(\x,q) :=
  \begin{cases} 
    0 & \text{ if } \x+\c_q \text{ is gas,}\\
    \frac{1}{2}[\varphi(\x+\c_q) + \varphi(\x)] & \text{ if } \x+\c_q \text{ is interface, }\\
    1 & \text{ if } \x+\c_q \text{ is liquid, }
  \end{cases}
  \label{eq:advection2}
\end{equation}
\end{subequations}
which means that there is no mass flux to gas cells. 

\begin{figure}
  \centering
  \caption{\label{fig:anisotropic}Simulation of a droplet splashing
    onto a liquid film \emph{without} surface tension. Reynolds number
    $\approx 250$.}
  \begin{subfigure}[t]{\linewidth}
    \includegraphics[width=0.49\linewidth]{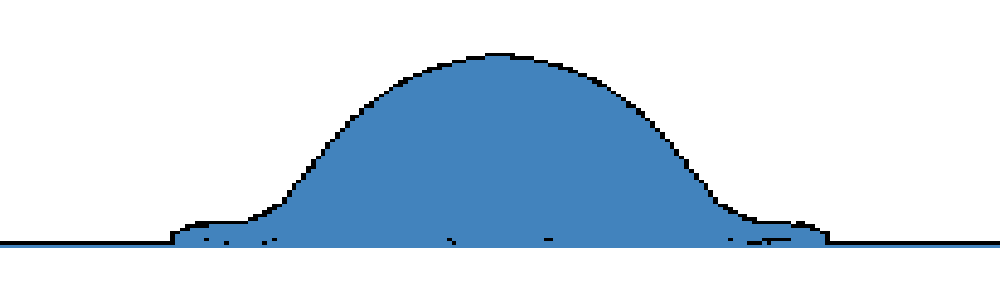}
    \includegraphics[width=0.49\linewidth]{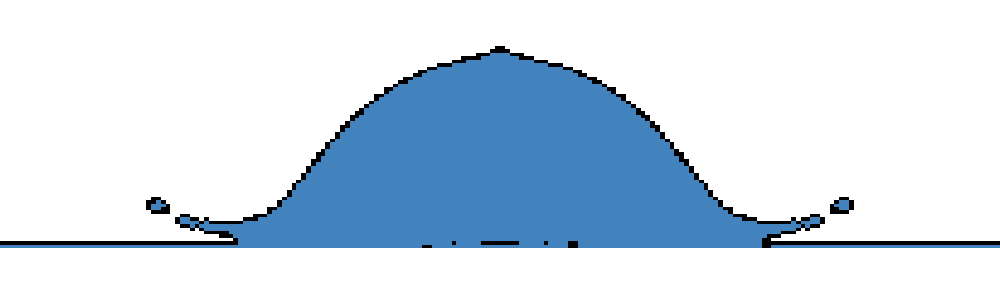}
    \caption{Slice through domain ($x=y$). Left: Imposing the free
      boundary condition on all links $\c_q \cdot \vec{n} \leq 0$ does
      not reproduce the rim correctly. Right: Rim clearly visible with
      the present implementation.}
  \end{subfigure}
  \begin{subfigure}[t]{\linewidth}
    \includegraphics[width=0.49\linewidth]{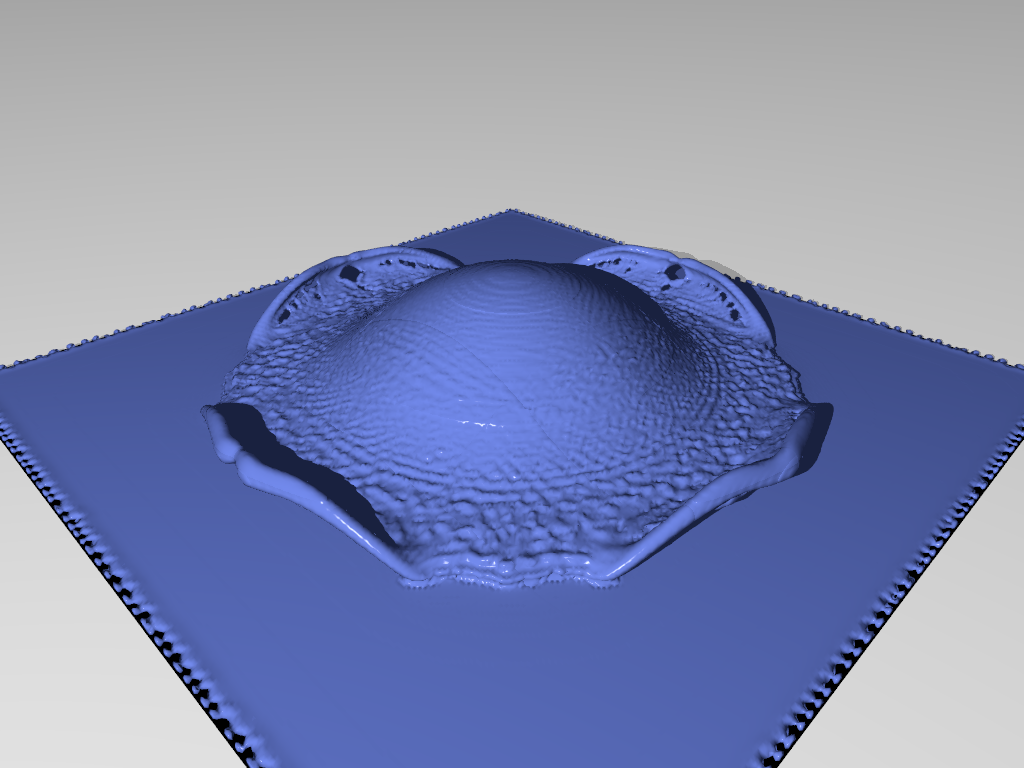}
    \includegraphics[width=0.49\linewidth]{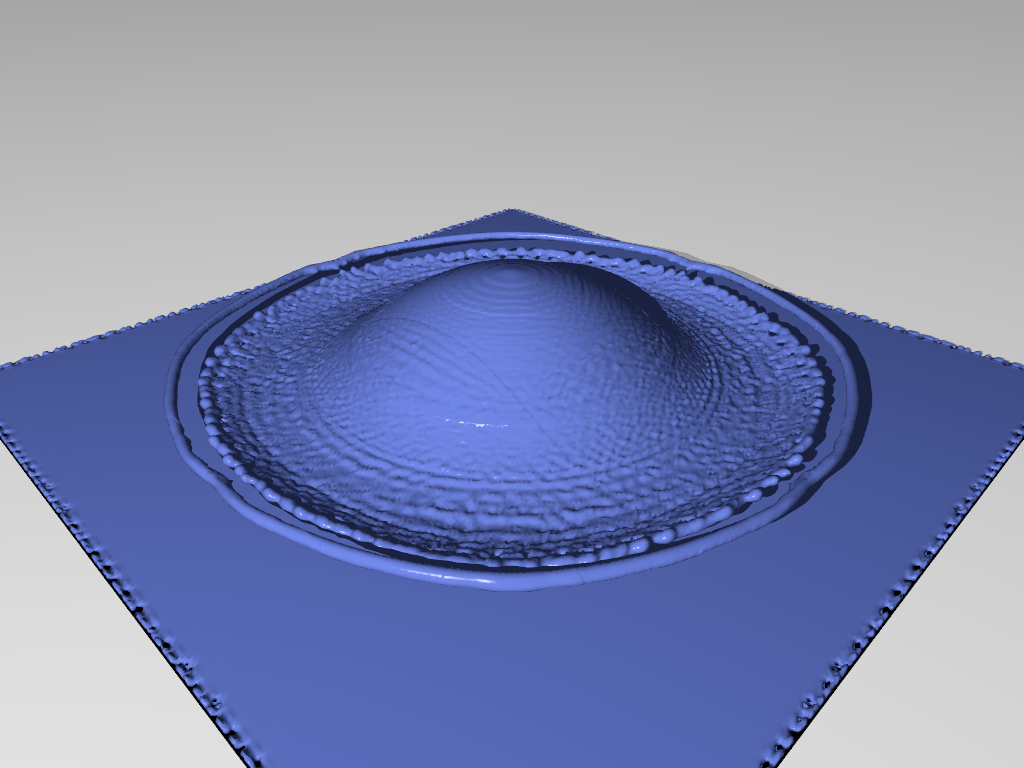}
    \caption{Iso-contour of the tracked interface. Left: Imposing the
      free boundary condition on all links $\c_q \cdot \vec{n} \leq 0$
      leads to incorrect crown formation. Right: Same simulation based
      on present implementation.}
  \end{subfigure}

\end{figure}

The free surface boundary condition for the LBM at the interface cells
is
\begin{equation}
  f_{\bar{q}}(\vec{x}, t+1) = f^{eq}_{q}(\rho_b, \vec{u}_b) + f^{eq}_{\bar{q}}(\rho_b, \vec{u}_b) - f'_{q}(\vec{x}, t),
  \label{eq:reconstruct}
\end{equation}
for all directions $q$ pointing towards gas cells $\vec{x}+\c_q \notin
\Omega(t)$. Hereby, $f_q'$ denotes the post-collision distribution
oriented towards the gas phase and $f^{eq}_{q}(\rho_b, \vec{u}_b)$ is
the equilibrium distribution function \cite{KoernerEtAl}. It can be
shown that Eq.~\eqref{eq:reconstruct} approximates a free boundary
condition with first order spatial accuracy \cite{Bogner2015}. The two
boundary condition parameters $\rho_b$ and $\vec{u}_b$ are the
macroscopic pressure and velocity at the interface, respectively. 
The boundary value $\rho_b$ is defined as
\begin{equation}
  \rho_b = \frac{1}{c_s^2} \left( p_g +  p_s \right),
\end{equation}
where $p_g(t)$ is the static pressure at the free surface and
$p_s(\vec{x}, t)$ is the Laplace pressure. The latter depends on the
local curvature of the interface by
\begin{equation}
  p_s(\vec{x}) = 2 \sigma \kappa(\vec{x}),
\end{equation}
with a constant surface tension $\sigma$.  We remark, that in the
original work of \cite{KoernerEtAl}, Eq.~\eqref{eq:reconstruct} is
applied for all $q$, oriented outwards with respect to the interface,
i.e., $\vec{c}_q \cdot \vec{n} \leq 0$, where $\vec{n} = \nabla
\varphi$ is a local normal vector to the interface pointing towards
the liquid (cf. Sec.~\ref{sec:FD}). However, we find that this
approach leads to anisotropic artifacts in the free surface
dynamics. Figure~\ref{fig:anisotropic} documents one according
example.
Hence, for the present work, we take into account the interface
orientation only at the contact line (\emph{corner nodes}) where the
boundary becomes inhomogeneous. As shown in Fig.~\ref{fig:inhomo}, a
contact line cell has both links to solid wall (off-boundary, S) nodes
and to gas (off-boundary, G) nodes. Links to solid wall nodes
(off-boundary nodes S) are usually subject to a no-slip condition
\footnote{We use the bounce back rule to realize no-slip boundary
  conditions, which assumes the wall position half-way between cell
  centers.}. The present implementation replaces the no-slip condition
with the free surface condition in corner nodes, if the outgoing
lattice direction has negative orientation with respect to the
inward-oriented interface normal $\vec{n}$ projected into the solid
wall.

\begin{figure}
  \centering
  \begin{tikzpicture}[scale=1.5]
    \fill[lightgray] (1,0) rectangle (2,2);
    \fill[lightgray] (2,1) rectangle (3,2);
    \draw[help lines] (0,0) grid (3,2);
    
    \node (7) at (0.5, -0.5) {S};
    \node (8) at (1.5, -0.5) {S};
    \node (9) at (2.5, -0.5) {S};
    
    \node (4) at (0.5, 0.5) {G};
    \node (5) at (1.5, 0.5) {I};
    \node (6) at (2.5, 0.5) {L};

    \node (1) at (0.5, 1.5) {G};
    \node (2) at (1.5, 1.5) {I};
    \node (3) at (2.5, 1.5) {I};

    \draw[black,->] (5) to (1);
    \draw[->] (5) to (4);
    \draw[thick,line width=1.5,->] (5) to (7);
    \draw[->] (5) to (8);
    \draw[->] (5) to (9);

    \draw[red,thick,line width=1.5,->] (5) to (2.0, -0.5);
    \node (n) at (2.0, -0.7) {$\color{red}{\vec{n}}$};

    \draw[thick] (0,0) -- (3,0);
    \fill[pattern=north east lines, thin] (0,-0.12) rectangle (3,0);

    \node (p) at (3.1,1.8) [label=right:\underline{Node types:} ] {};
    \node (p) at (3.1,1.5) [label=right:L: liquid] {};
    \node (p) at (3.1,1.3) [label=right:I: interface] {};
    \node (p) at (3.1,1.1) [label=right:G: gas (off-boundary)] {};
    \node (p) at (3.1,0.9) [label=right:S: solid (off-boundary)] {};
  \end{tikzpicture}
  \caption{\label{fig:inhomo}Inhomogeneous boundary at an interface
    corner node with boundary-intersecting lattice directions
    indicated by arrows. The free boundary rule is imposed on the
    thick arrow direction due to its orientation with respect to the
    interface with normal $\vec{n}$.}
\end{figure}

 The remaining part of this
section lays out the three different approaches considered in this
study, to extract the local interface curvature $\kappa$ from the volume fraction $\varphi$. Notice
that the notion of a lattice \emph{cell} reflects the represented
cubic volume with the length of one grid spacing, and which is used to
define the fill levels. However, the LBM data is more precisely
located at lattice \emph{nodes} which coincide with the centers of the
cells.

\subsection{\label{sec:FD}Finite Difference Approximation (FD)}
As first proposed in \cite{Brackbill92}, one way to approximate the
curvature $\kappa$ of the boundary surface defined through the fill
levels, is to compute a finite difference approximation to
\begin{equation}
  \kappa = - (\nabla \cdot \hat{\vec{n}}),
  \label{eq:curvature}
\end{equation}
where $\hat{\vec{n}}$ is the normalized gradient of the indicator
function.
To obtain this gradient, we use central finite differences to
approximate
\begin{equation}
  \vec{n} = \nabla \varphi,
  \label{eq:normal}
\end{equation}
which can be interpreted also as a local normal vector to the free
surface. We use the finite difference approximation suggested in
\cite{ParkerYoungs}. However, the curvature computation according to
Eq.~\eqref{eq:curvature} means that second order derivatives of the
\emph{non-smooth} indicator function $\varphi$ are approximated by
finite differences, which inevitably introduces larger errors. Hence,
much work has been published (cf. for instance \cite{Williams98}) on
effective ways to mollify the fill level information and smooth out
surface tension in the ``continuum surface tension'' approach.  We use
the $K_8$ - kernel with support radius $\epsilon=2.0$ (cf.
\cite{Williams98}), i.e., only the next neighbor information is
included in the convolution.

Wetting properties are included by directly specifying an ideal
equilibrium contact angle $\theta_{eq}$ for solid boundaries. For
obstacle cells with surface normal $\hat{\vec{n}}_w$ the boundary
condition at the solid wall is
\begin{equation}
  \hat{\vec{n}} = \hat{\vec{n}}_w \cos{\theta_{eq}} + \hat{\vec{n}}_t \sin{\theta_{eq}},
  \label{eq:contactNormal}
\end{equation}
where $\hat{\vec{n}}_t$ is a tangent vector to the wall and normal to
the contact line \cite{Brackbill92}. Since the wall position rarely
coincides with the lattice nodes, the boundary value according to
Eq.~(\ref{eq:contactNormal}) is extrapolated to the obstacle
node. Hereby, the vector $\hat{\vec{n}}_t$ is computed by projection
of the interface normal at the fluid boundary cell $\vec{x}_b$ onto
the wall. The boundary condition for the fill level in the obstacle
cells (affecting the interface normal $\vec{n}(\vec{x}_b)$ in the
boundary cells) is a reflection condition for the
$\varphi$-values. Here, we generally compute the boundary value for
the obstacle cells $\x_o \notin \Omega$ based on the neighboring inner
nodes $\x_o+\c_q \in \Omega$ using the formula
\begin{equation}
  \varphi(\x_o) =  \sum_{|\hat{\vec{n}}_w \cdot \hat{\vec{c}}_q| > \alpha \atop \x_o+\c_q \in \Omega } |\hat{\vec{n}}_w \cdot \hat{\vec{c}}_q| \varphi( \x_o+\c_q )  \biggl/  \sum_{ |\hat{\vec{n}}_w \cdot \hat{\vec{c}}_q| > \alpha \atop \x_o+\c_q \in \Omega } |\hat{\vec{n}}_w \cdot \hat{\vec{c}}_q|,
\end{equation}
with an apperture $\alpha=\sqrt{2}/2$, to achieve a smoothed
reflection for the boundary values. 

\subsection{\label{sec:TR}Triangular Reconstruction (TR) based on piecewise linear interface construction (PLIC)}
In \cite{Pohl2007}, a curvature computation is suggested based on a
local triangulation of interface points in a $3 \times 3 \times 3$
neighborhood around each interface cell. The interface points are
determined using a \emph{piecewise linear interface construction} (PLIC)
approach \cite{ScardovelliZaleski}. For an interface cell centered
around $\vec{x}_i$, let $P$ be the half space $P = \{\vec{x} |
(\vec{x} - \vec{p}) \cdot \vec{n}(\vec{x}_i) \geq 0 \}$, where
$\vec{p} = \vec{x}_i + a \vec{n}(\vec{x}_i)$ is a point within the
corresponding unit volume $V(\vec{x}_i)$. Now, the interface point
$\vec{p}$ is defined such that the cut-off volume $V \cap P$ satisfies
\begin{equation}
  vol(V \cap P) = \varphi(\vec{x}_i).
\end{equation}
We determine $a$ iteratively, similar to \cite{Rider98}, with an error
bound of $\approx 4\times 10^{-13}$ (40 iterations in a bisection
algorithm) assuming an exact surface normal. The interface point can
be computed in one step together with the estimation of the surface
normals. For the latter we employ again a finite difference scheme to
Eq.~\eqref{eq:normal}, however, without any convolution
step. Alternative, higher order PLIC algorithms are discussed in
\cite{Pilliod2004,Tryggvason2001}.

Once the interface points are determined by the PLIC scheme, the
algorithm described in \cite{Pohl2007} is used to construct a local
``triangle fan'' from the interface points within the local
neighborhood. Then, a variant of the algorithm described in
\cite{Taubin95} determines the curvature of this polygonal surface.
Notice, that the described TR scheme as well as the curvature
estimation by LSQR of Sec.~\ref{sec:LSQR} are based solely on the
surface points and use the gradient information represented by the
interface normals only as far as it is needed to construct these
interface points. The TR method can be extended to support adhesive
boundary conditions. In \cite{Donath2011}, a way to extend the local
triangulations at solid boundaries to achieve an ``artificial
curvature'' matching with a desired equilibrium contact angle, is
described. However, the implementation of the geometric construction
is difficult in three dimensions. Also, we found the results obtained
from that method often not convincing, which motivated the least
squares - based approach of the following section.

\subsection{\label{sec:LSQR}Least Squares Reconstruction (LSQR) based
  on piecewise linear interface construction (PLIC)}
\begin{figure}
  \begin{tikzpicture}[scale=1.5]
    \draw[help lines] (0,0) grid (3,2);
    \draw[thick] (0,0) -- (3,0);
    \fill[pattern=north east lines, line width=0] (0,-0.12) rectangle (3,0);
    \draw[blue, domain=0.9:3] plot (\x, {1.3*sqrt(\x-0.9)});
    
    \node (p) at (0.6,0.8) [label=right:$\vec{p}_i$] {};
    \node (cp) at (0.6,0) [label=above:$\vec{p}_{s,i}$] {};
    \node (n) at (1.3, 0.6) [label=right:$\color{red}{\vec{n}_i}$] {};
    \node (nw) at (0.6, -0.3) [label=right:$\color{red}{\vec{n}_w}$] {};
    
    \draw[red, dashed, thick, domain=0.5:2.1] plot (\x, {0.71+(\x-1.2)*(1.187)});
    \draw[red,thick,->] (1.2,0.71) to (1.4, 0.45);
    \draw[red,thick,->] (0.6,0.0) to (0.6,-0.4);
    \fill[black] (1.2,0.71) circle (1.7pt);
    \fill[black] (0.6,0) circle (1.7pt);
  \end{tikzpicture}
  \caption{\label{fig:contactPoint}Determination of contact point
    $\vec{p}_{s,i}$ for a contact line cell. The PLIC segment defined
    by interface point $\vec{p}_i$ and interface normal $\vec{n}_i$ is
    extended and intersected with the obstacle wall.}
  %
    

\end{figure}
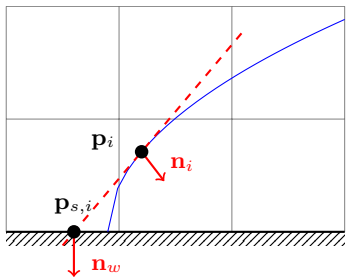
The third approach to include surface tension consists in
reconstructing the interface as a quadratic function in each $3 \times
3 \times 3$ neighborhood around an interface cell. It has previously
been described in \cite{Popinet2009}, there however without the
inclusion of wall adhesion effects. Like the TR approach, it is based
on the PLIC of interface points described above in
Sec.~\ref{sec:TR}. Let $\hat{\vec{t}}_u$, $\hat{\vec{t}}_v$ and
$\hat{\vec{n}}$ be a local orthonormal basis, i.e., $\hat{\vec{t}}_u$
$\hat{\vec{t}}_v$ tangential to the interface, and $\vec{p}$ the local
interface point. Now, assume that $i$ is indexing all the remaining
interface cells in a $3 \times 3 \times 3$ - neighborhood. The
interface cell data $(\hat{\vec{n}}_i, \vec{p}_i)$ is used to fit the
model function
\begin{subequations}
\begin{equation}
  f(u,v) = A u^2 + B v^2 + C u v + H u + I v + J
  \label{eq:LSQR-model}
\end{equation}
with parameters $(A, B, C, H, I, J)$, by minimizing the error
\begin{equation}
  E = \sum_i{ | f( u_i , v_i  ) - f_i  |^2 }.
  \label{eq:LSQR-error}
\end{equation}
\end{subequations}
Here, $u_i = (\vec{p}_i-\vec{p}) \cdot \hat{\vec{t}}_u$, $v_i =
(\vec{p}_i-\vec{p}) \cdot \hat{\vec{t}}_v$, and $f_i =
(\vec{p}_i-\vec{p}) \cdot \hat{\vec{n}}$. This yields a linear least
squares problem that has to be solved locally. We obtain the best
results, when fixing the constant parameter $J=0$, i.e., accepting
only solutions that interpolate the surface point $\vec{p}$ of the
respective interface cell. We use the implementation from the LAPACK
library \cite{lapack} based on QR decomposition of the corresponding
system matrix. Once $f$ is determined the curvature can be evaluated
analytically, using
\begin{equation}
  \kappa = \frac{ A (1+I^2) + B (1+H^2) - 2 C H I}{(\sqrt{1 + H^2 + I^2})^3}.
  \label{eq:anacurv}
\end{equation}
The approach can be seen as a modified version of the PROST - scheme
from \cite{Renardy2002}. Both schemes fit a parabolic function in a
local neighborhood around each interface cell. However, as a major
difference, PROST fits $f(x,y,z)$ directly to the fill levels
minimizing the error $\sum_i{ ( V_i(f) - \varphi_i )^2 }$ of the
cut-off volumes $V_i$ that the iso-surface $f(x,y,z)=0.5$ cuts out of
the interface cell $i$. This makes the least squares problem
non-linear and $f$ has to be determined by iteratively computing the
error and updating of coefficients. The scheme described here is
computationally less expensive.

To include the effect of boundary adhesion, we extend the method in
the following way: If the local $3 \times 3 \times 3$ -neighborhood
contains an obstacle cell, then for each contact-line cell (i.e., an
interface cell that has an obstacle cell as neighbor) from the same
neighborhood, one contact point is approximated with a contact normal
$\hat{\vec{m}}$ defined according to Eq.~(\ref{eq:contactNormal}). In
the contact point, we require
\begin{equation}
  \nabla f(u_c, v_c) = - (m_u, m_v)/m_n,
  \label{eq:contactConstraint}
\end{equation}
where $m_u = \hat{\vec{m}} \cdot \hat{\vec{t}}_u$, $m_v =
\hat{\vec{m}} \cdot \hat{\vec{t}}_v$, and $m_n = \hat{\vec{m}} \cdot
\hat{\vec{n}}_u$, and $u_c$, $v_c$ are the coordinates of the contact
point in the locally defined tangential plane.  If the neighboring
interface cell is the center of the current $3 \times 3 \times 3$ -
neighborhood, we use Eq.~\ref{eq:contactConstraint} as a constraint to
the respective optimization problem given by the
Eqs.~(\ref{eq:LSQR-model}-\ref{eq:LSQR-error}).  Otherwise, the
condition is simply included in the optimization of the error,
Eq.~\eqref{eq:LSQR-error}.  
To obtain the contact point, we construct the closest intersection of the
interface segment of the contact line cell with the solid surface as in
Fig.~\ref{fig:contactPoint}.

\subsection{\label{sec:limitations}Current limitations}
The interface tracking of the present implementation defines
\emph{interface cells} as active lattice Boltzmann cells that have a
D3Q19 neighborhood containing both gas and liquid cells. This
definition turned out to impose a limitation when simulating thin
liquid films on wetting surfaces and with contact angles below
$45^\circ$. As shown in Fig.~\ref{fig:limitation}, the thickness of a
liquid film on solid substrate has to be resolved at least by one
liquid cell in height, for the method to work correctly. A film
thickness smaller than 1 lattice unit is not supported. For strongly
wetting surfaces ($\theta_{eq}<45^{\circ}$), we often observe
anisotropic errors because it is then problematic to impose the
correct contact angles according to the LSQR method
(Sec.~\ref{sec:LSQR}) and the TR method (Sec.~\ref{sec:TR}): Depending
on the approximated interface position represented by the fill level
information, the computed curvature values then tend to oscillate and
overshoot. In Fig.~\ref{fig:valid}, for instance, the approximation of
the interface as a smooth surface through the two leftmost interface
cells can be expected erroneous under the condition of an acute
intersection angle ($\theta_{eq}<45^\circ$) with the solid surface. It
is important to notice that these restrictions are specific to the
presented FSLBM algorithm, while the presented curvature
reconstruction schemes can be applied in any VOF-context.



\begin{figure}
  \centering
  \caption{\label{fig:limitation}The bottom-left interface (I) cell in Fig.~\ref{fig:invalid} has several gas (G) neighbors but no liquid neighbor (L). Hence, the configuration of Fig.~\ref{fig:invalid} is not supported by the present implementation. Shown in Fig.~\ref{fig:valid} is a valid configuration since all interface cells have both liquid and gas neighbors. The minimum supported film thickness is therefore at least one (liquid) lattice cell.}
  \begin{subfigure}[t]{0.48\linewidth}
    \begin{tikzpicture}
      \fill[lightgray] (1,0) rectangle (3,1);
      \fill[lightgray] (2,1) rectangle (4,2);
      \fill[lightgray] (3,2) rectangle (4,3);
      \draw[help lines] (0,0) grid (4,3);
      \draw[blue, domain=1.4:4] plot (\x, {6/5*(\x-1.4)-3/25*(\x-1.4)*(\x-1.4)});
      
      \node at (0.5, 0.5) {G};
      \node at (1.5, 0.5) {I};
      \node at (2.5, 0.5) {I};
      \node at (3.5, 0.5) {L};

      \node at (0.5, 1.5) {G};
      \node at (1.5, 1.5) {G};
      \node at (2.5, 1.5) {I};
      \node at (3.5, 1.5) {I};

      \node at (0.5, 2.5) {G};
      \node at (1.5, 2.5) {G};
      \node at (2.5, 2.5) {G};
      \node at (3.5, 2.5) {I};

      \draw[thick] (0,0) -- (4,0);
      \fill[pattern=north east lines, line width=0] (0,-0.12) rectangle (4,0);
    \end{tikzpicture}
    \caption{\label{fig:invalid}invalid.}
  \end{subfigure}
  \begin{subfigure}[t]{0.48\linewidth}
    \begin{tikzpicture}
      \fill[lightgray] (1,0) rectangle (2,2);
      \fill[lightgray] (2,1) rectangle (3,3);
      \fill[lightgray] (3,2) rectangle (4,3);
      \draw[help lines] (0,0) grid (4,3);
      \draw[blue, domain=0.6:4] plot (\x, {6/5*(\x-0.6)-3/25*(\x-0.6)*(\x-0.6)});
      \node at (0.5, 0.5) {G};
      \node at (1.5, 0.5) {I};
      \node at (2.5, 0.5) {L};
      \node at (3.5, 0.5) {L};

      \node at (0.5, 1.5) {G};
      \node at (1.5, 1.5) {I};
      \node at (2.5, 1.5) {I};
      \node at (3.5, 1.5) {L};

      \node at (0.5, 2.5) {G};
      \node at (1.5, 2.5) {G};
      \node at (2.5, 2.5) {I};
      \node at (3.5, 2.5) {I};

      \draw[thick] (0,0) -- (4,0);
      \fill[pattern=north east lines, line width=0] (0,-0.12) rectangle (4,0);
    \end{tikzpicture}
    \caption{\label{fig:valid}valid.}
  \end{subfigure}
\end{figure}


\section{\label{sec:results}Numerical Results}
\definecolor{bluegrey}{RGB}{100,100,200}

The numerical results presented in the following have been obtained
with the waLBerla lattice Boltzmann framework \cite{walberla2011} that
includes the FSLBM implementation described in \cite{Pohl2007}.

\subsection{Equilibrium Spherical Bubble}
\label{sec:equilibriumBubble}
\begin{figure}
  \centering
  \includegraphics[width=\linewidth]{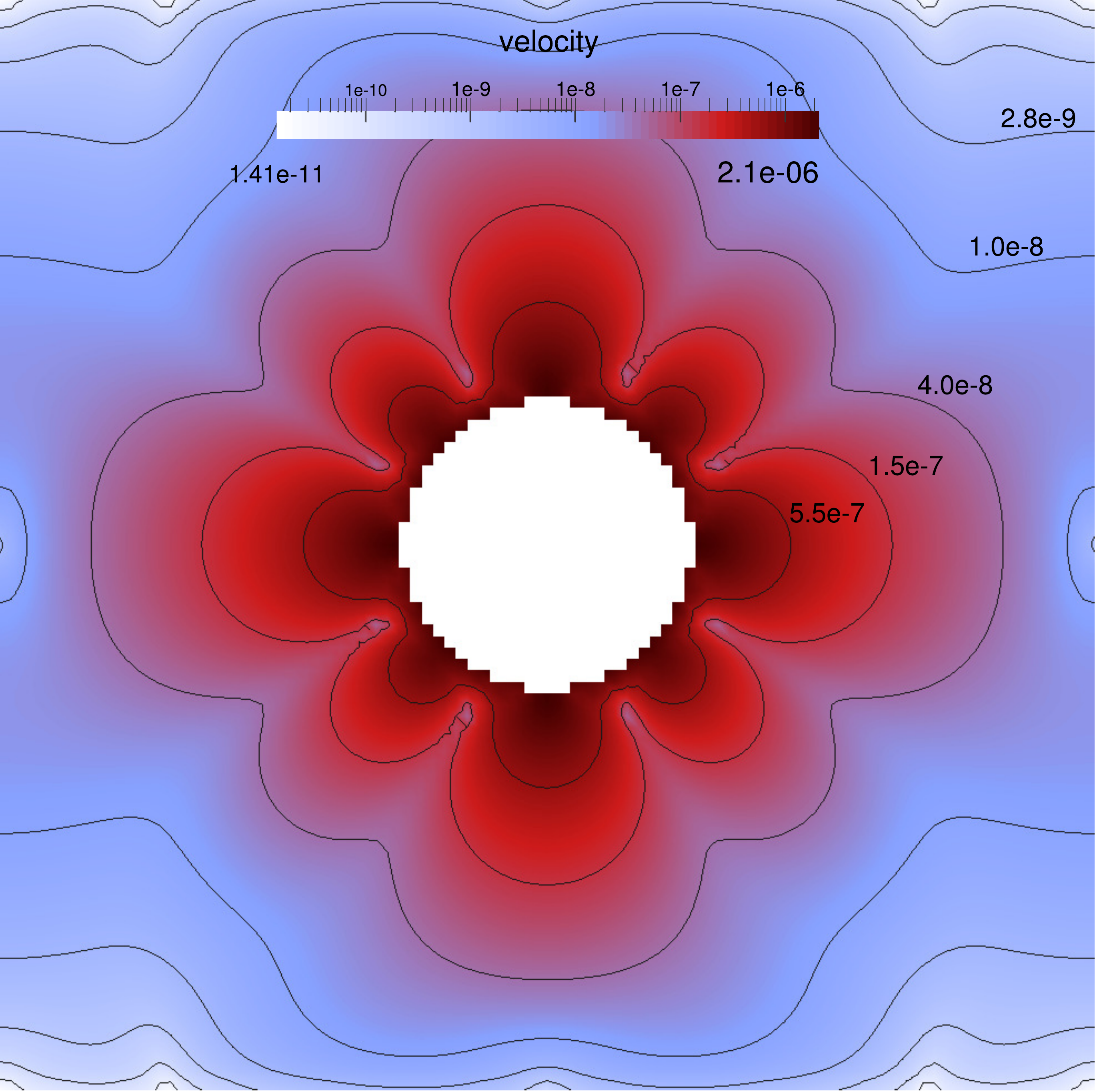}
  \caption{Slice through a domain containing a single gas
    bubble. The color indicates the magnitude (lattice units) of the
    spurious currents due to errors in the curvature estimation with
    the FD-approach.}
  \label{fig:spurious}
\end{figure}

The standard benchmark for surface tension models is a static
equilibrium bubble. If the curvature estimation would return the exact
value $\kappa = 1/R$ everywhere on the interface, the solution to the
problem would be a perfectly vanishing velocity field. Due to the
existing errors, however, spurious currents occur around the interface
from regions with overestimated Laplace pressure to positions
underestimating the value (cf. Fig.~\ref{fig:spurious})
\cite{ScardovelliZaleski}. For non-sophisticated methods, the
magnitude of the spurious velocities can be related to the ratio
$\sigma / \mu$, of surface tension and dynamic viscosity $\mu=\rho
\nu$, with a constant prefactor $\leq 1$. This means that the error is
independent of the spatial resolution or converging very slowly, and
thus poses a limitation in terms of applicability to problems
involving dominant capillary forces. See
\cite{Harvie2005,Williams98,ZaleskiEtAl2009} for a discussion of the
problem in connection with the VOF method.

Here we adopt the problem stated in \cite{Renardy2002} using
non-dimensional values with respect to a cubic domain of unit length,
containing a spherical bubble of radius $R=0.125$ centered at $(0.5,
0.5, 0.5)$. We apply no-slip boundary conditions at top ($z=1$) and
bottom $z=0$ sides of the domain and periodicity along all other
directions. The viscosity of the liquid of density $\rho=4$ is set to
$\mu=1$, and the surface tension parameter is $\sigma=0.357$.  The
dimensionless Ohnesorge number ($Oh = \mu / \sqrt{\sigma \rho R}$)
corresponding to the given problem is $Oh \approx 2.37$. Notice that
in \cite{Renardy2002} the bubble is actually a second fluid of the
same density and viscosity as the liquid (two-phase), while in our
case the flow inside the gas bubble is not simulated but represented
only in terms of a gas pressure value that is dynamically adjusted
according to the changes of the gas volume upon interface advection
(free surface flow with bubble model
\cite{Anderl2014a,Cobussat2005}). We remark that the present test case
is appropriate to evaluate the error in the curvature computation
only. Since there is no flow velocity in this test case aside from the
spurious currents, one expects no significant error contribution from
the advection of the indicator function or the LBM. We have therefore
evaluated the test case first in the standard case with a static frame
of reference (Sec.~\ref{sec:LaplaceStatic}), and second in a moving
frame of reference (Sec.~\ref{sec:LaplaceMoving}) with a constant
uniform background velocity added to the flow.

\subsubsection{\label{sec:LaplaceStatic}Static frame of reference}

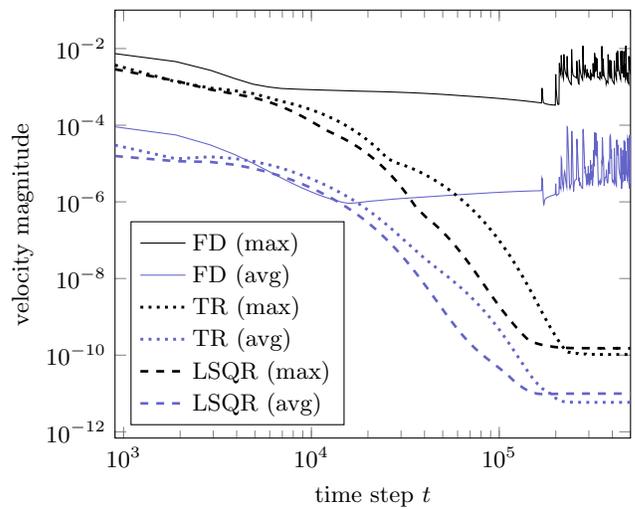
\begin{figure}
  \centering
  %

  \begin{tikzpicture}
    \begin{loglogaxis}[
      xlabel=time step $t$,
      xmin=900,xmax=500000,
      ylabel=velocity magnitude,
      legend cell align=left,
      legend pos=south west
      ]
      \addplot[color=black,mark=''] table[x=time, y=max_FD] {./SpuriousCurrentsDecayTrimmed.csv};
      \addlegendentry{FD (max)}
      \addplot[color=bluegrey,mark=''] table[x=time, y=avg_FD] {./SpuriousCurrentsDecayTrimmed.csv};
      \addlegendentry{FD (avg)}

      \addplot[color=black,dotted,mark='',line width=1.1] table[x=time, y=max_TR] {./SpuriousCurrentsDecayTrimmed.csv};
      \addlegendentry{TR (max)}
      \addplot[color=bluegrey,dotted,mark='',line width=1.1] table[x=time, y=avg_TR] {./SpuriousCurrentsDecayTrimmed.csv};
      \addlegendentry{TR (avg)}

      \addplot[color=black,dashed,mark='',line width=1.0] table[x=time, y=max_LSQR] {./SpuriousCurrentsDecayTrimmed.csv};
      \addlegendentry{LSQR (max)}
      \addplot[color=bluegrey,dashed,mark='',line width=1.0] table[x=time, y=avg_LSQR] {./SpuriousCurrentsDecayTrimmed.csv};
      \addlegendentry{LSQR (avg)}
    \end{loglogaxis}
  \end{tikzpicture}
  \caption{Temporal evolution of maximal and average (spurious) velocity for the
    different surface tension models at a fixed grid spacing
    $\delta_x=1/96$. After the decay of an initial shock, the magnitude of the
    spurious currents in the system can be evaluated. The FD model shows the
    largest errors and often leads to oscillative behaviour. Similar behavior is
    obtained for $\delta_x = 1/48$, $1/144$, and $1/192$.}
  \label{fig:spuriousDecay}
\end{figure}

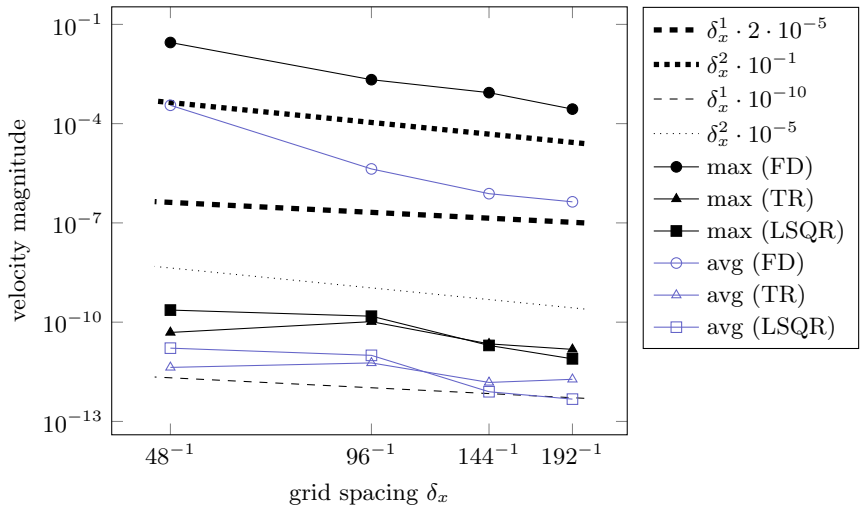
\begin{figure*}
  \begin{tikzpicture}[trim axis left]
    \begin{loglogaxis}[
      xlabel=grid spacing $\delta_x$,x dir=reverse,
      xticklabels={$48^{-1}$, $96^{-1}$, $144^{-1}$, $192^{-1}$},
      xtick={0.0208333333, 0.0104166667, 0.0069444444, 0.0052083333},
      ylabel=velocity magnitude,
      legend cell align=left,
      legend pos=outer north east
      ]
      \addplot[line width=2,black,dashed,domain=0.005:0.022] {2e-5*x }; \addlegendentry{$ \delta_x^1 \cdot 2 \cdot 10^{-5}$}
      \addplot[line width=2,black,dotted,domain=0.005:0.022] {1*x^2 }; \addlegendentry{$\delta_x^2 \cdot 10^{-1}$}

      \addplot[dashed,domain=0.005:0.022] {1e-10*x }; \addlegendentry{$ \delta_x^1 \cdot 10^{-10}$}
      \addplot[dotted,domain=0.005:0.022] {1e-5*x^2 }; \addlegendentry{$\delta_x^2 \cdot 10^{-5}$}

      \addplot[color=black,mark=*] table[x=dx, y=error] {
        dx              error
        0.0208333333	0.028325
        0.0104166667	0.0021301458
        0.0069444444	0.0008639653
        0.0052083333	0.0002749667

      };
      \addlegendentry{max (FD)}

      \addplot[color=black,mark=triangle*] table[x=dx, y=error] {
        dx              error
        0.0208333333	4.886375E-011
        0.0104166667	1.03624583333333E-010
        0.0069444444	2.20902777777778E-011
        0.0052083333	1.5055625E-011

      };
      \addlegendentry{max (TR)}

      \addplot[color=black,mark=square*] table[x=dx, y=error] {
        dx              error
        0.0208333333	2.3191875E-010
        0.0104166667	1.51490625E-010
        0.0069444444	1.98145138888889E-011
        0.0052083333	7.79395833333333E-012

      };
      \addlegendentry{max (LSQR)}

      \addplot[color=bluegrey,mark=o] table[x=dx, y=error] {
        dx              error
        0.0208333333	0.0003585375
        0.0104166667	4.27811458333333E-006
        0.0069444444	7.64131944444444E-007
        0.0052083333	4.334734375E-007
      };
      \addlegendentry{avg (FD)}

      \addplot[color=bluegrey,mark=triangle] table[x=dx, y=error] {
        dx              error
        0.0208333333	4.29270833333333E-012
        0.0104166667	5.86341666666667E-012
        0.0069444444	1.50839583333333E-012
        0.0052083333	1.87496875E-012
      };
      \addlegendentry{avg (TR)}

      \addplot[color=bluegrey,mark=square] table[x=dx, y=error] {
        dx              error
        0.0208333333	1.63838541666667E-011
        0.0104166667	9.93127083333333E-012
        0.0069444444	7.88409722222222E-013
        0.0052083333	4.72019270833333E-013
      };
      \addlegendentry{avg (LSQR)}
    \end{loglogaxis}
  \end{tikzpicture}
  \caption{\label{fig:spuriousConvergence}Dependency of spurious
    velocity (maximum and average) on grid spacing $\delta_x$ for the
    three different models evaluated after $500,000$ time steps. The
    dotted and dashed lines represent first and second order slopes,
    respectively. For the FD scheme, the plot indicates a first order
    convergence for both maximum and average spurious flow speed. The
    errors of the FD scheme are orders of magnitude above those of the
    methods based on surface reconstruction.}
\end{figure*}
We perform a resolution study varying the grid spacing between the
values $\delta_x = 1/48, 1/96, 1/144, 1/192 $ at a fixed time step of
$\delta_t = 10^{-4}$, which yields the resolution-dependent lattice
relaxation times $\tau = 0.6728$, $1.191$, $2.055$ and $3.264$. The
surface tension parameter in lattice units varies accordingly and
takes the values $\sigma_L/10^{-3} =0.0987$, $0.7896$, $2.665$ and
$6.317$. For initialization of the fill levels at $t=0$ with the
spherical geometry, we employ a spatial subdivision technique,
refining each discrete cell volume by a factor of 100 along each
coordinate. The simulation then exhibits a series of pressure
disturbances until the numerical equilibrium is
reached. Fig.~\ref{fig:spuriousDecay} shows the development of the
maximal and average flow velocity within the domain for $500,000$ time
steps. After a certain number of time steps the shock wave is
sufficiently decayed for both the TR and the LSQR method, such that
both maximal and average flow velocity within the domain become
smaller than $\lesssim 10^{-10}$ in magnitude. The FD approach,
however does not converge and enters into an oscillating behavior
instead. Here, the spurious currents are large enough to trigger
changes in the layer of interface nodes. This also introduces sudden
changes in the curvature computation, thus explaining the
oscillations. Fig.~\ref{fig:spuriousConvergence} shows the maximal and
average velocity within the domain for different spatial resolutions
and various methods. The strength of the spurious currents obtained
with the FD method are in accordance with the values reported in
\cite{Renardy2002}, where a similar approach (CSF) is used for
referencing. The curvature information obtained by geometric
reconstruction (TR and LSQR methods) is much more accurate than the FD
approximation, and reduces the spurious velocities almost down to the
order of machine precision.

We also evaluate the accuracy of the curvature values obtained for the
numerical equilibrium, i.e., the state reached after $500,000$ time
steps. Fig.~\ref{fig:curvatureError} compares the error in curvature
at various grid spacings for the three different methods. Only the
plot for the LSQR method indicates a second order rate of
convergence. However, it is clear that the occurrence of spurious
currents is not due to constant over- or underestimation of curvature,
but rather because of its variance with the node position. This
becomes obvious, when comparing Fig.~\ref{fig:curvatureStdev} showing the
standard deviation over all interface nodes in the final state to the
resulting spurious currents of Fig.~\ref{fig:spuriousConvergence}.

\begin{figure*}
  \caption{Comparison of curvature estimation for a stationary bubble
    after $500,000$ time steps. For the FD-model, the included graphs
    are somewhat arbitrary, because the values oscillate over time,
    analog to the spurious currents in
    (cf. Fig.~\ref{fig:spuriousDecay}). }
  \begin{subfigure}[t]{0.48\linewidth}
    \begin{tikzpicture}[trim axis left]
      \begin{loglogaxis}[
        xlabel=grid spacing $\delta_x$,x dir=reverse,
        xticklabels={$48^{-1}$, $96^{-1}$, $144^{-1}$, $192^{-1}$},
        xtick={0.0208333333, 0.0104166667, 0.0069444444, 0.0052083333},
        ylabel=$L^2(\kappa)$,
        legend cell align=left,
        legend pos=south west,legend style={at={(-0.0,-0.0)}}
        ]
        \addplot[dashed,domain=0.005:0.025] {x }; \addlegendentry{$\sim \delta_x^1$}
        \addplot[dotted,domain=0.005:0.025] {30*x^2 }; \addlegendentry{$\sim \delta_x^2$}

        \addplot[color=red,mark=*] table[x=dx, y=error] {
          dx              error
          0.0208333333	0.0902448
          0.0104166667	0.0490684
          0.0069444444	0.016254
          0.0052083333	0.0155673
        };
        \addlegendentry{FD}

        \addplot[color=black,mark=square*] table[x=dx, y=error] {
          dx              error
          0.0208333333	0.0180768
          0.0104166667	0.00421798
          0.0069444444	0.00177479
          0.0052083333	0.000991554
        };
        \addlegendentry{LSQR}

        \addplot[color=blue,mark=triangle*] table[x=dx, y=error] {
          dx              error
          0.0208333333	0.00773397
          0.0104166667	0.00617829
          0.0069444444	0.00519107
          0.0052083333	0.00517137
        };
        \addlegendentry{TR}

      \end{loglogaxis}
    \end{tikzpicture}
    \caption{\label{fig:curvatureError}The $L^2$ norm of the curvature
      error. Only the LSQR curvature error does converge in the test,
      with a rate of convergence in $\mathcal{O}(\delta_x^2)$. }
  \end{subfigure}
  \begin{subfigure}[t]{0.48\linewidth}
    \begin{tikzpicture}[trim axis right]
      \begin{loglogaxis}[
        xlabel=grid spacing $\delta_x$,x dir=reverse,
        xticklabels={$48^{-1}$, $96^{-1}$, $144^{-1}$, $192^{-1}$},
        xtick={0.0208333333, 0.0104166667, 0.0069444444, 0.0052083333},
        ylabel=$STDEV(\bar{\kappa})$,
        ytick pos=right,ylabel near ticks,
        legend style={at={(0.03,0.5)},anchor=west}
        ]
        \addplot[color=red,mark=*] table[x=dx, y=stdev] {
          dx              stdev
          0.0208333333	1.0521024
          0.0104166667	0.343152
          0.0069444444	0.127167408
          0.0052083333	0.123288192
        };
        \addlegendentry{FD}

        \addplot[color=black,mark=square*] table[x=dx, y=stdev] {
          dx              stdev
          0.0208333333	2.31408E-007
          0.0104166667	3.1994304E-007
          0.0069444444	8.449776E-008
          0.0052083333	8.448192E-008
        };
        \addlegendentry{LSQR}

        \addplot[color=blue,mark=triangle*] table[x=dx, y=stdev] {
          dx              stdev
          0.0208333333	2.765856E-007
          0.0104166667	0.000000171
          0.0069444444	1.19405232E-007
          0.0052083333	0.000000169

        };
        \addlegendentry{TR}
      \end{loglogaxis}
    \end{tikzpicture}
    \caption{\label{fig:curvatureStdev}Standard deviation of average
      curvature values. The FD scheme has the largest standard
      deviation. Consecutively, the FD scheme generates the largest
      spurious currents.}
  \end{subfigure}
\end{figure*}

\subsubsection{\label{sec:LaplaceMoving}Moving frame of reference}
\begin{figure}
  \centering
  %
  \begin{tikzpicture}
    \begin{axis}[
      xlabel=time $t/T$,
      xmin=0,xmax=1.0,
      ylabel=$L^2(\kappa)$,ymin=0,ymax=1.2,
      legend cell align=left,
      legend style={at={(0.8,0.55)}}
      ]
      \addplot[color=red,mark=''] table[x=time, y=FD] {./PROST-moving-curvature.csv};
      \addlegendentry{FD}
      \addplot[color=blue,mark=''] table[x=time, y=TR] {./PROST-moving-curvature.csv};
      \addlegendentry{TR}
      \addplot[color=black,mark=''] table[x=time, y=LSQR] {./PROST-moving-curvature.csv};
      \addlegendentry{LSQR}
    \end{axis}
  \end{tikzpicture}    
  \caption{\label{fig:ProstMoving}$L^2$ - error in curvature for the
    three different surface tension models for a moving frame of
    reference. Simulation of a spherical bubble during traversal a
    periodic domain with uniform velocity. At $t=0$ the bubble is
    initialized to a nearly ideal (spherical) shape. Due to errors in
    the advection operator and the surface tension models the error
    increases and after $t=0.4$ oscillates around a fixed value. The
    FD scheme is less sensitive to errors in the fill levels,
    presumably because it is more diffusive and based on the mollified
    indicator function. }
\end{figure}
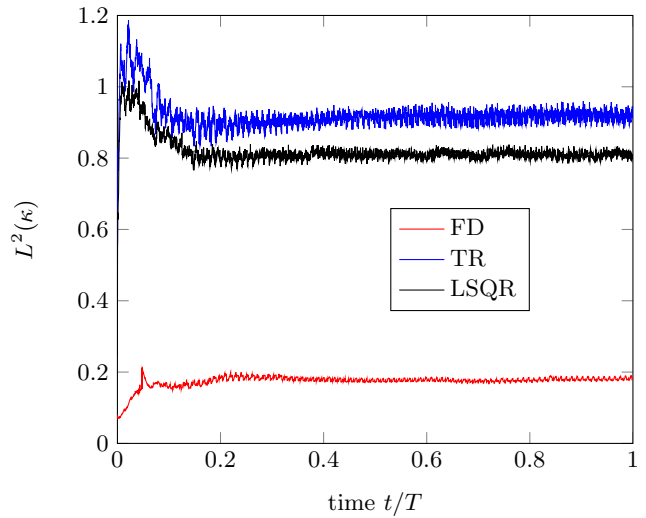

\begin{figure}
  \centering
  \includegraphics[width=\columnwidth]{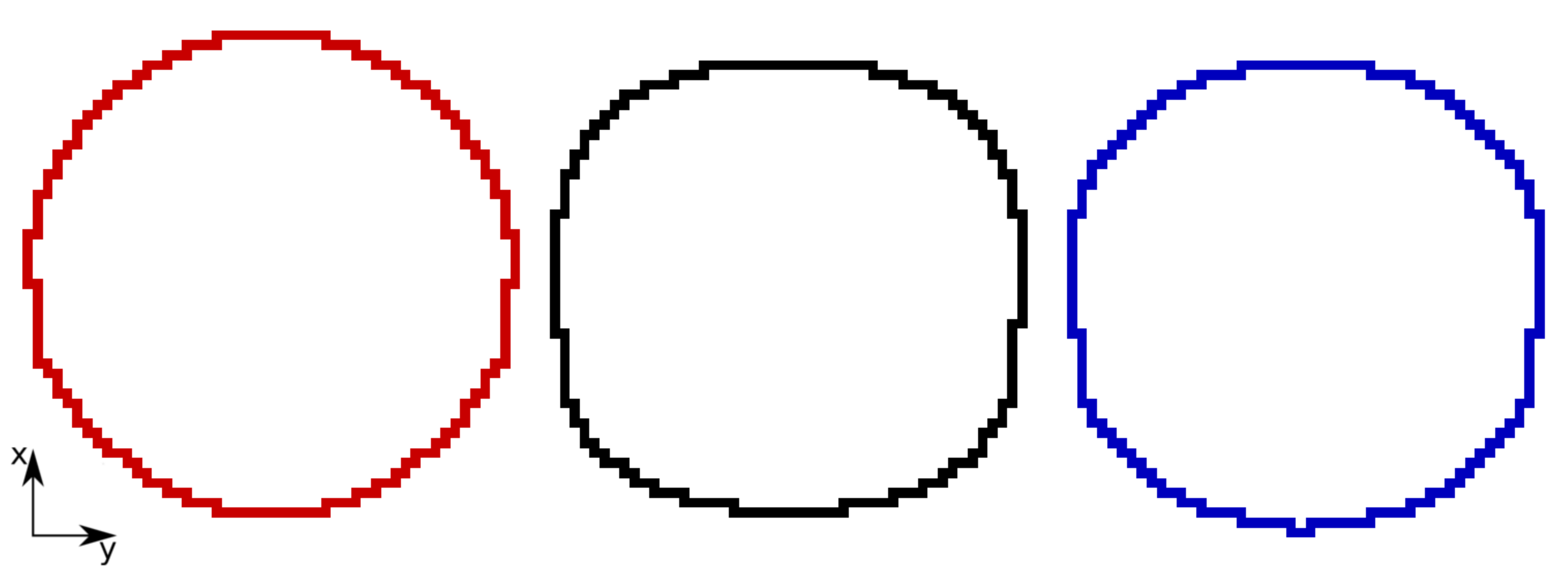}
  \caption{\label{fig:MovingShapes}Comparison of the shapes of the
    bubbles after advection along $x$-axis with surface tension. From
    left to right: FD, LSQR, TR. Visualization shows the layer of
    interface cells in the $x$-$y$-plane through the center of the
    bubbles at time $0.94 T$ ($150000$ time steps). The initial radius
    of the bubble was $24 \delta_x$.}
\end{figure}

When dealing with dynamical problems numerical errors in the advection
of the indicator function $\varphi$ are often critical. This holds in
particular for the present test case, since any errors in the fill
levels will introduce an additional error into the computed curvature
values. To study effects of advection, we add a constant background
velocity of $\vec{u}_{0}=(1, 0, 0)^T$ to the test case. This means
that there is now a constant advection involved and the spurious
currents appear as a deviation from the background velocity. For
$\delta_x = 1/96$, $\delta_t=2.5\times10^{-5}$ (lattice relaxation
time $\tau=0.6728$, $u_{x}=0.0024 \delta_x/\delta_t$), we run the
simulation for $T = L/u_{0,x}$ time steps, i.e., the bubble traverses
the periodic domain of length $L$ exactly one time. This can be
interpreted as a moving frame of reference while, physically, the
setup is equivalent to the static version of
Sec.~\ref{sec:LaplaceStatic}. Measuring the curvature error over time,
Fig.~\ref{fig:ProstMoving} exhibits a dramatic increase in error as
compared to the static test case. The errors of the reconstructive
methods, LSQR and TR, are now significantly larger than the error of
the FD model. A possible reason is that the FD model is more diffusive
than the reconstruction methods, and thus less sensitive to errors in
the indicator function field. A grid study with space steps $\delta_x
= 1/48, 1/96, 1/144, 1/192$ and corresponding time steps $\delta_t =
1\times10^{-4}, 2.5\times10^{-5}, 1.11\times10^{-5},
6.25\times10^{-6}$ (diffusive scaling) revealed that these errors do
not converge with the grid spacing. Figure~\ref{fig:MovingShapes}
shows that the shape of the bubble after advection deviates notably
from a true sphere. In accordance with the increased errors in
curvature, the reconstruction based schemes show the most deviation.
Also, in this case the spurious currents do no longer converge for
either method. This indicates that there is an additional error that
stems from the advection of the indicator function $\varphi$.

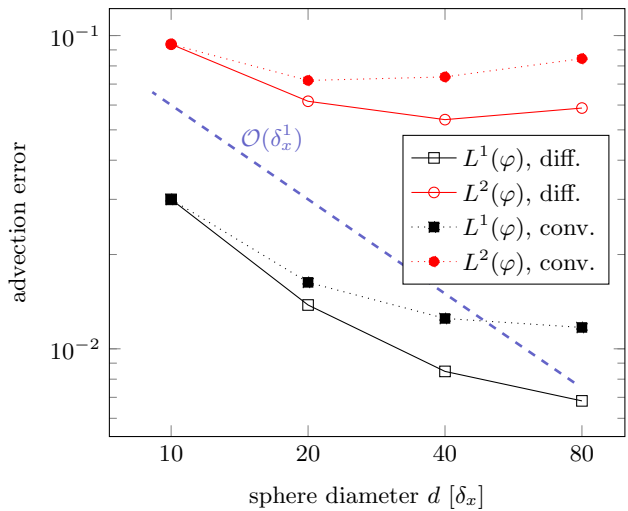
\begin{figure}
  \centering
  \begin{tikzpicture}[trim axis left]
    \begin{loglogaxis}[
      xlabel={sphere diameter $d$ $[\delta_x]$},x dir=reverse,
      xticklabels={$10$, $20$, $40$, $80$},
      xtick={1, 0.5, 0.25, 0.125},
      ylabel=advection error,
      legend cell align=left,
      legend style={at={(0.97,0.53)},anchor=east}
      ]
      \addplot[mark=square] table[x=dx, y=error] {
        dx       error
        1	3.00E-02
        0.5	1.38E-02
        0.25	8.47E-03
        0.125	6.82E-03
        
      };
      \addlegendentry{$L^1(\varphi)$, diff.}
      \addplot[red,mark=o] table[x=dx, y=error] {
        dx       error
        1	9.38E-02
        0.5	6.17E-02
        0.25	5.39E-02
        0.125	5.87E-02
      };
      \addlegendentry{$L^2(\varphi)$, diff.}

      \addplot[dotted,mark=square*] table[x=dx, y=error] {
        dx       error
        1	3.00E-02
        0.5	1.63E-02
        0.25	1.25E-02
        0.125	1.17E-02
      };
      \addlegendentry{$L^1(\varphi)$, conv.}
      \addplot[dotted,red,mark=*] table[x=dx, y=error] {
        dx       error
        1	9.38E-02
        0.5	7.19E-02
        0.25	7.38E-02
        0.125	8.44E-02
      };
      \addlegendentry{$L^2(\varphi)$, conv.}

      \addplot[color=bluegrey,line width=1.0,dashed,domain=0.13:1.1] {0.06*x}; 
      \node [above] at (axis cs:  0.6, 0.04) {$\color{bluegrey}{\mathcal{O}(\delta_x^1)}$};
    \end{loglogaxis}
  \end{tikzpicture}
  \caption{\label{fig:advection}Dependency of $L^1$ and $L^2$ - errors in the
    indicator function $\varphi$ on lattice resolution under convective and
    diffusive scaling (cf. Tab.~\ref{tab:advection}). Test case consisted of a
    spherical bubble of diameter $d$ is advected over a distance $d$. The
    indicated order of convergence is below~1.}
\end{figure}

Since we are not aware of any numerical evaluation of the FSLBM
advection scheme, we supplement the Laplace test with a convergence
check of the indicator function values. To this end, a gas bubble of
diameter $d=10 \delta_x$ is placed inside of a periodic computational
domain $\Gamma$ with a prescribed uniform velocity $\vec{u}_0 = (0.05,
0, 0)^{T} \delta_x/\delta_t$. The LBM data is thus constant with
$\vec{f}(\x,t) = \vec{f}^{eq}(p_{g}/c_s^2,\vec{u_0})$, where $p_g$ is
the constant reference pressure. This excludes any error contribution
by the LBM or the surface tension modeling to the indicator function
$\varphi$. From the prescribed LBM data we compute the advection of
the bubble in terms of the indicator function according to
Eqs.~\eqref{eq:advection1} and \eqref{eq:advection2}.  After a time
$T=d/u_{0,x}$ the bubble has moved a distance equal to its
diameter. To evaluate the error, we use the $L^1$ and $L^2$ error
norms, by comparing the fill levels at time $T$ to the initial
configuration $t=0$, i.e.,
\begin{subequations}
  \begin{align}
    L^1(\varphi) &= \frac{\sum_{\x\in\Gamma} |\varphi(\x + T \vec{u}_0 , T) - \varphi(\x,0)| }{\sum_{\x\in\Gamma} |\varphi(\x,0)|},\\
    L^2(\varphi) &= \sqrt{\frac{\sum_{\x\in\Gamma} [\varphi(\x + T \vec{u}_0 , T) - \varphi(\x,0)]^2 }{\sum_{\x\in\Gamma}\varphi(\x,0)^2}}.
\end{align}
\end{subequations}
The test case is repeated with successively refined grid spacing,
under convective scaling ($\delta_t \sim \delta_x$) and diffusive
scaling ($\delta_t \sim \delta_x^2$). The exact parameterizations and
the resulting errors are collected in
Tab.~\ref{tab:advection}. Fig.~\ref{fig:advection} shows dependency of
the respective errors on the lattice resolution. The indicated
convergence rate is below first order, independent of the used scaling
and error norm. This means that one cannot assert first order
convergence with the present advection scheme in a simple translation
test. 

With respect to the moving Laplace bubble test, the increased error
observed in Fig.~\ref{fig:ProstMoving} as compared to the static one
presented in Sec.~\ref{sec:LaplaceStatic}, as well as the degenerated
bubble shapes after advection (cf. Fig.~\ref{fig:MovingShapes})
suggests the following explanation. Any errors in the indicator
function affect also the curvature computation, which explains the
temporal oscillation of the error as the bubble moves relative to the
grid (cf. Fig.~\ref{fig:ProstMoving}). The reconstruction methods (TR
and LSQR) seem more sensitive to the advective errors than the simpler
FD scheme, and hence are less stable in the dynamic case. Even though
the curvature estimates of LSQR and TR are more accurate and
effectively reduce spurious currents in the static benchmarks, the
combination with the low-order advection scheme is problematic.

\begin{table*}
  \centering
  \begin{tabular}{cc||cccc|cccc}
    & & \multicolumn{4}{c}{convective} & \multicolumn{4}{c}{diffusive} \\
    $d/\delta_x$ & $\delta_x$ & $\delta_t$ & $u_{0,x} \delta_t/\delta_x$ & $L^1(\varphi)$ & $L^2(\varphi)$ & $\delta_t$ & $u_{0,x} \delta_t/\delta_x$ & $L^1(\varphi)$ & $L^2(\varphi)$\\ 
    \hline 
    $10$ & 1   & 1   & $0.05$ & 3.00\% & 9.38\% & 1    & $0.05$    & 3.00\% &9.38\%\\ 
    $20$ & 1/2 & 1/2 & $0.05$ & 1.63\% & 7.19\% & 1/4  & $0.025$   & 1.38\% &6.17\%\\ 
    $40$ & 1/4 & 1/4 & $0.05$ & 1.25\% & 7.38\% & 1/16 & $0.0125$  & 0.85\% &5.39\%\\ 
    $80$ & 1/8 & 1/8 & $0.05$ & 1.17\% & 8.44\% & 1/64 & $0.00625$ & 0.68\% &5.87\%\\ 
  \end{tabular}
  \caption{\label{tab:advection}$L^1$ and $L^2$ - errors in the indicator function $\varphi$ due to advection. The translation of a spherical bubble of diameter $d$ over a distance $d$ in uniform velocity field (constant velocity $\vec{u}_0$ along $x$-axis) is evaluated for different time and grid spacings $\delta_t$ and $\delta_x$. The test case has been performed for both convective and diffusive scaling.}
\end{table*}

\subsection{\label{sec:bcWetting}Droplets on wetting boundaries}
\subsubsection{\label{sec:equilibrium}Equilibrium sessile droplets}
Next, we evaluate the error at the contact line with solid
boundaries. We change the setup of Sec.~\ref{sec:equilibriumBubble} to
a spherical cap shaped droplet, such that the initial state of the
droplet is close to the ideal equilibrium. The equilibrium is a
spherical cap resting on the wall, with a sphere radius $R$ related to
the equilibrium contact angle $\theta_{eq}$ of the wall by $R = h /
(1-\cos{\theta_{eq}})$. Given the volume of the droplet $V$, the ideal
equilibrium height of the droplet is
\begin{equation}
  h = \sqrt[3]{ \frac{V}{\pi (\frac{1}{1-\cos{\theta_{eq}}} - \frac{1}{3} )} },
\end{equation}
and the ideal contact line radius $r$ (base radius of the spherical
cap) is
\begin{equation}
  r = \sqrt{ \frac{1}{3} \left( \frac{6 V}{\pi h} - h^2 \right) }.
\end{equation}
The \emph{simulated} height $h^*$ and contact line radius $r^*$ have
to be \emph{approximated} from the indicator function, as
\begin{subequations}
\begin{equation}
  h^* = \max_{\x \in I}( x_z + \varphi(\x) -0.5),
\end{equation}
and
\begin{equation}
  r^* = \frac{1}{|I_c|} \sum_{\x  \in I_c} \tilde{r}(\x),
\end{equation}
where $I$ denotes the set of interface nodes, $I \supset I_c$ the set
of contact line nodes. Hereby, $\tilde{r}$ is the local approximation
\begin{equation}
  \tilde{r}(\x) = \|\x-\x_o\| + (\varphi(\x)-0.5), 
\end{equation}
\end{subequations}
where $\x_o$ is the ideal center of the circular contact line. This
approximation reflects that the nodes $\x \in I$ are the centers of
the corresponding interface cells. Furthermore, the error in the
contact line is evaluated using the $L^2$ error definition,
\begin{equation}
  L^2(r) = \sqrt{\frac{\sum_{\x \in I_c} [r - \tilde{r}(\x)]^2}{\sum_{\x \in I_c} r^2 }}.
\end{equation}

\paragraph{The error analysis} involved the ideal contact angles
$\theta_{eq} = 30^\circ$, $45^\circ$, $60^\circ$, $90^\circ$,
$120^\circ$, $135^\circ$ and $150^\circ$ at a constant resolution with
$\delta_x=1/96$ and $\delta_t=10^{-4}$.  In all cases, the droplet
volume $V$ was chosen equal to that of a hemisphere of radius
$R=0.125$. Fig.~\ref{fig:sessileDroplet} shows the relative errors in
height of the droplet shape obtained in the simulations, and the
relative $L^2$ error in the simulated contact line radius. For the
most extreme contact angles $\theta_{eq}$, the droplet height and
contact line move away significantly from the initial (ideal)
equilibrium position, increasing the respective errors until the
numerical equilibrium is reached. Fig.~\ref{fig:clRadius} also shows
the standard deviation in the measured contact line radius, computed
over all contact line cells. While the errors in the contact line seem
to be comparable in size, the higher values in $STDEV(r^*)$ obtained
with the FD scheme indicate that the simulated contact lines are more
anisotropic than with the LSQR scheme.


\begin{figure*}
  \caption{\label{fig:sessileDroplet}Sessile droplet equilibrium on
    solid surface for various equilibrium contact angles $\theta_{eq}$
    simulated with grid spacing $\delta_x=1/96$. Deviation of
    numerical equilibrium from ideal equilibrium for FD and LSQR
    scheme.}
  \begin{subfigure}[t]{0.48\linewidth}
    \begin{tikzpicture}[trim axis left]
      \begin{axis}[
        xlabel=contact angle $\theta_{eq}$,
        ylabel=error $h^*/h-1.0$,
        legend cell align=left,
        legend pos=north east
        ]
        \addplot[color=red,mark=*] table[x=Theta, y=Heighterrorbrackbill] {stationaryDroplet.csv};
        \addlegendentry{FD}

        \addplot[color=black,mark=square*] table[x=Theta, y=Heighterrorparabolic] {stationaryDroplet.csv};
        \addlegendentry{LSQR}
      \end{axis}
    \end{tikzpicture}
    \caption{The relative deviation of the simulated droplet height $h^*$.}
  \end{subfigure}
  \hspace{0.0cm}
  \begin{subfigure}[t]{0.48\linewidth}
    \begin{tikzpicture}[trim axis right]
      \begin{axis}[
        xlabel=contact angle $\theta_{eq}$,
        ylabel=error / standard deviation,ytick pos=right,ylabel near ticks, 
        legend cell align=left,
        legend pos=north west
        ]
        \addplot[color=black,mark=square*] table[x=Theta, y=CLL2errorparabolic] {stationaryDroplet.csv};
        \addlegendentry{$L^2(r)$, LSQR}
        \addplot[color=red,mark=*] table[x=Theta, y=CLL2errorbrackbill] {stationaryDroplet.csv};
        \addlegendentry{$L^2(r)$, FD}

        \addplot[dashed,color=black,mark=square*] table[x=Theta, y=CLStdevparabolic] {stationaryDroplet.csv};
        \addlegendentry{$STDEV(r^*)$, LSQR}
        \addplot[dashed,color=red,mark=*] table[x=Theta, y=CLStdevbrackbill] {stationaryDroplet.csv};
        \addlegendentry{$STDEV(r^*)$, FD}

      \end{axis}
    \end{tikzpicture}
    \caption{\label{fig:clRadius}The relative deviation of contact
      line radius in the $L^2$ norm, and the standard deviation of
      $r^*$ computed over all contact line cells as a measure for
      anisotropic artifacts.}
  \end{subfigure}
\end{figure*}

\begin{figure*}
  \centering
  \caption{\label{fig:DropletConvergence}Grid convergence study of the
    $L^2$ error in the simulated contact line radius - FD approach and
    LSQR approach in comparison.}

  \begin{subfigure}[b]{0.48\linewidth}
    \begin{tikzpicture}[trim axis left]
      \begin{loglogaxis}[
        xlabel=space step $\delta_x$,x dir=reverse,
        xticklabels={$48^{-1}$, $96^{-1}$, $144^{-1}$},
        xtick={0.0208333333, 0.0104166667, 0.0069444444},
        ylabel=$L^2(r)$,ymax=0.15,ymin=0.002,
        legend cell align=left,
        legend pos=south west
        ]
        \addplot[color=bluegrey,line width=1.0,dashed,domain=0.006:0.025] {4.0*x}; \addlegendentry{$\sim \delta_x^1$}
        
        \addplot[dashed,mark=*] table[x=dx, y=error] {
          dx       error
          0.0208333333   0.0610761	
          0.0104166667   0.0256658
          0.0069444444   0.0207583
        };
        \addlegendentry{$\theta_{eq}=60^\circ$}
        
        \addplot[mark=*] table[x=dx, y=error] {
          dx       error
          0.0208333333  0.0426878
          0.0104166667  0.0190497
          0.0069444444  0.0134927
        };
        \addlegendentry{$\theta_{eq}=90^\circ$}

        \addplot[dotted,mark=*] table[x=dx, y=error] {
          dx       error
          0.0208333333  0.0973444
          0.0104166667  0.0389345
          0.0069444444  0.0387821
        };
        \addlegendentry{$\theta_{eq}=120^\circ$}

      \end{loglogaxis}
    \end{tikzpicture}
    \caption{FD scheme.}
  \end{subfigure}
  \begin{subfigure}[b]{0.48\linewidth}
    
    \begin{tikzpicture}[trim axis right]
      \begin{loglogaxis}[
        xlabel=space step $\delta_x$,x dir=reverse,
        xticklabels={$48^{-1}$, $96^{-1}$, $144^{-1}$},
        xtick={0.0208333333, 0.0104166667, 0.0069444444},
        ylabel=$L^2(r)$,ymax=0.15,ymin=0.002,
        ytick pos=right,ylabel near ticks,
        legend pos=north east,
        legend cell align=left,
        ]

        \addplot[dashed,mark=square*] table[x=dx, y=error] {
          dx       error
          0.0208333333   0.0594971
          0.0104166667   0.0211923
          0.0069444444   0.0134882
        };
        \addlegendentry{$\theta_{eq}=60^\circ$}

        \addplot[mark=square*] table[x=dx, y=error] {
          dx       error
          0.0208333333  0.00559683	
          0.0104166667  0.00508519
          0.0069444444  0.0029987
        };
        \addlegendentry{$\theta_{eq}=90^\circ$}

        \addplot[dotted,mark=square*] table[x=dx, y=error] {
          dx       error
          0.0208333333  0.0597827	
          0.0104166667  0.0322758
          0.0069444444  0.0248789
        };
        \addlegendentry{$\theta_{eq}=120^\circ$}

        \addplot[color=bluegrey,line width=1.0,dashed,domain=0.006:0.025] {x}; 
      \end{loglogaxis}
    \end{tikzpicture}
    \caption{LSQR scheme.}
  \end{subfigure}
\end{figure*}
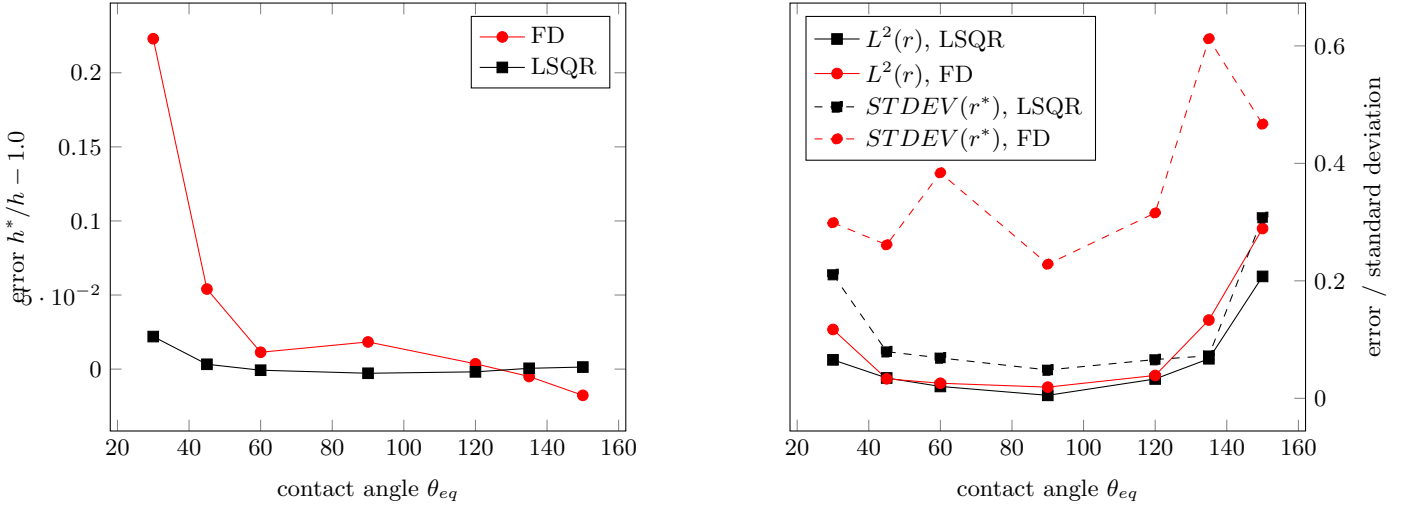

\paragraph{A convergence study} was performed for the selected
equilibrium positions of $\theta_{eq} = 60^\circ$, $90^\circ$ and
$120^\circ$ by altering the grid spacing to $\delta_x = 1/48$, $1/96$
and $1/144$, using diffusive scaling for the time step.
Fig.~\ref{fig:DropletConvergence} compares the $L^2$ errors obtained
by the FD and the LSQR scheme and shows that the LSQR error is
generally smaller. The error behavior in terms of grid dependency is
somewhat irregular for both schemes. However, at least for the LSQR
model the convergence rate of the error appears to be approximately
first order. Since the construction of the contact points described in
Sec.~\ref{sec:LSQR} exploits the \emph{linear} approximation of the
reconstructed interface, the observed first order error is in
accordance with the expected behavior.

\subsubsection{Droplet spreading on wetting boundaries}
Because of the limitations described in Sec.~\ref{sec:limitations},
for extreme contact angles, $\theta_{eq}<45^\circ$ or
$\theta_{eq}>135^\circ$, we focus our study to a few dynamical cases.
For the inertial regime (low Ohnesorge number), a spherical droplet in
contact with an adhesive plane substrate, will start to spread
according to the power law
\begin{equation}
  r(t) \sim t^{0.5},
  \label{eq:powerLaw}
\end{equation}
where $r$ is the radius of the circular contact line
\cite{Biance2004}. A numerical simulation of contact line dynamics
requires the accurate modeling of a slip condition to resolve the
stress singularities in the moving contact line. Furthermore, it is in
general not sufficient to work with a static contact angle model that
imposes the equilibrium contact angle $\theta_{eq}$ everywhere at the
contact line \cite{Sui2014}. In the presented FSLBM, the interface
representation by volume fractions introduces a certain amount of
\emph{numerical} slip that allows the free surface to move in the
contact line \cite{Renardy2001}. This numerical slip is related to the
grid spacing and does not necessarily recapture the correct
physics. We have not introduced a dynamic contact angle model. 

\begin{figure}
  \centering

  %
  \begin{tikzpicture}[trim axis left]
    \begin{loglogaxis}[
      xlabel=time step $t$,xmin=300,xmax=6000,
      xticklabels={$300$, $500$, $1000$, $2000$, $4000$},
      xtick={300, 500, 1000, 2000, 4000},
      yticklabels={$4$, $8$, $16$},
      ytick={4, 8, 16},
      ylabel=radius $r^*$,
      legend pos=south east,
      legend cell align=left,
      ]
      \addplot[color=red,dashed,line width=1.0,mark=''] table[x=time, y=FD_10mPa] {./spreadingDynamicsTrimmed.csv};
      \addlegendentry{FD, $\nu_1$}
      \addplot[color=black,dashed,line width=1.0,mark=''] table[x=time, y=LSQR_10mPa] {./spreadingDynamicsTrimmed.csv};
      \addlegendentry{LSQR, $\nu_1$}

      \addplot[color=red,mark=''] table[x=time, y=FD_50mPa] {./spreadingDynamicsTrimmed.csv};
      \addlegendentry{FD, $\nu_2$}
      \addplot[color=black,mark=''] table[x=time, y=LSQR_50mPa] {./spreadingDynamicsTrimmed.csv};
      \addlegendentry{LSQR, $\nu_2$}

      \addplot[color=bluegrey,dashed,line width=1.2,domain=300:6000] {0.2*x^0.5}; 
      \addplot[color=bluegrey,dotted,line width=1.2,domain=300:6000] {0.9*x^0.35}; 
      \node [above] at (axis cs:  850, 10) {$\color{bluegrey}{\mathcal{O}(t^{0.35})}$};
      \node [above] at (axis cs:  1600, 6) {$\color{bluegrey}{\mathcal{O}(t^{0.5})}$};
    \end{loglogaxis}
  \end{tikzpicture}    
  \caption{\label{fig:dropletSpreading}Contact line dynamics of the
    spreading droplet; comparison of simulations based on the FD
    scheme and the LSQR scheme. The test case is repeated for two
    different viscosities, $\nu_1=3.32226\times10^{-3}$, and
    $\nu_1<\nu_2=1.66113\times10^{-2}$. Both schemes tend to
    underestimate the ideal spreading law of $r(t) \sim t^{0.5}$.}
\end{figure}
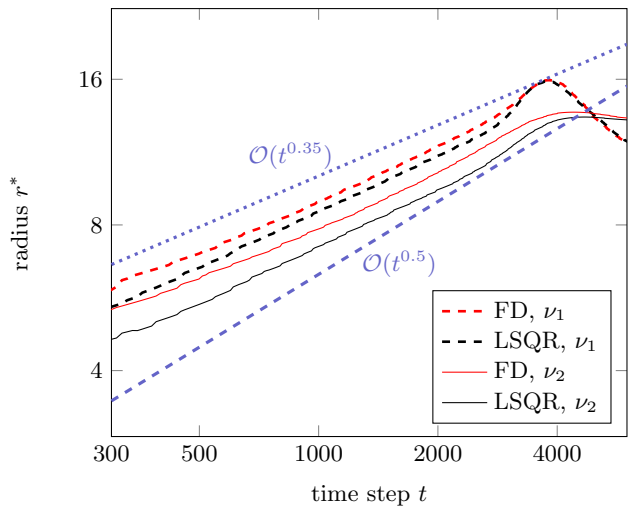

We simulate the spreading of a droplet of radius $R = 10$ lattice
units on a flat plate with ideal equilibrium contact angle
$\theta_{eq}=85^\circ$. The droplet is initialized as a sphere placed
at a distance of $R$ from the solid boundary. Due to a discretization
effect, the simulated initial contact line has a positive radius
$r>0$, such that the adhesive boundary condition can be imposed in the
contact line cells. The surface tension constant is chosen $\sigma =
4.3189\times10^4$ for a fluid of lattice reference density $\rho =
1.0$ and kinematic viscosity $\nu$ of $\nu_1 = 3.32226\times10^{-3}$
(first run, $\tau=0.509967$) and $\nu_2 = 1.66113\times10^{-2}$
(second run, $\tau=0.549834$). Fig.~\ref{fig:dropletSpreading} shows
the contact line radius of the simulation over time. Both the FD and
the LSQR model seem to recapture approximately
Eq.~\eqref{eq:powerLaw}, however, with a smaller exponent $<0.5$. This
is acceptable, considering the low grid resolution and the static
contact line model. The power law obtained, $r \sim t^{0.35}$, seems
similar to the one reported in \cite{Ding2007} for level set-based
simulations.


\section{\label{sec:conclusion}Conclusion}
Three different ways to compute the interface curvature from a VOF
indicator function have been realized for comparison within a free
surface LBM. A stationary Laplace bubble benchmark shows that methods
based on geometric reconstruction (TR and LSQR, in the present study)
can reach a significantly higher accuracy than \emph{continuum surface
  force}-like approaches that are based on finite difference
approximations (FD, in the present study). In accordance with previous
studies, reconstruction-based methods significantly reduce the
magnitude of spurious currents. The LSQR approach shows a second order
rate of convergence with respect to grid spacing, which could not be
achieved with the other two approaches.

For the generalized Laplace bubble test in a \emph{moving frame of
  reference}, a previous study \cite{Popinet2009} reports convergence
of errors, using a combination of higher order advection scheme and
curvature reconstruction in a Navier-Stokes discretization. This
behavior could not be reproduced with the present FSLBM. Our results
indicate that this lack of convergence is caused by the simplified
advection of the indicator function that does not take into account
any geometry information. In particular, the advection scheme fails to
converge in a simple uniform advection test, indicating a lower order
of accuracy than previously reported for simple reconstruction based
schemes (typically first or second order, with \emph{SLIC} or
\emph{PLIC}-based advection in \cite{Pilliod2004}). Not surprisingly, the
method thus fails to converge in the Laplace benchmark when conducted
in a moving frame of reference including advection of the
interface. This means that a conclusion drawn in \cite{DonathEtAl2010}
must be corrected: More accurate curvature estimation does not
necessarily improve the FSLBM in surface tension driven flows, since
(asymptotically) the dominant error is caused by the advection
scheme. Furthermore, it turned out that, in the moving case, the
curvature estimation by finite differences (FD) is less sensitive to
errors introduced by the advection scheme than the
reconstruction-based approaches (LSQR and TR).

We have successfully extended the LSQR model to adhesive boundaries,
and compared it to the FD model in several numerical test cases. The
order of convergence decreases to one in the presence of adhesive
boundaries. However, for the stationary case, the errors of the new
approach are still significantly smaller then those of obtained with
the non-reconstructive FD approach. In dynamic scenarios, like contact
line spreading on wetting surfaces, both models can recapture the
basic inertial power law dynamics. However, similar to the bulk
dynamic case, the error situation changes. Because the LSQR method is
more sensitive to errors stemming from the FSLBM advection than the FD
scheme, simulations do not profit from the higher-order scheme.

We conclude that a major improvement to the FSLBM would consist in a
replacement of the advection scheme. VOF advection based on geometric
reconstruction is significantly more complex and was, for this reason,
excluded from the present study. In principle, any known advection
scheme for VOF indicator functions
\cite{ScardovelliZaleski,Tryggvason2011} could be used to replace the
simplified advection scheme of the FSLBM
\cite{JanssenCAMWA2010}. Considering that the simulation of surface
tension according to the presented TR and LSQR schemes is based on a
PLIC scheme to reconstruct the interface geometry, switching to a
PLIC-based VOF advection scheme would be a possible solution. A major
benefit in accuracy can be expected.




\section*{Acknowledgments}
Parts of the presented results have been obtained during a stay at the Technical
University Eindhoven. The first author would like to thank the graduate school
of the cluster of excellence ``Engineering of Advanced Materials (EAM)'', and
the ``3TU Research Centre Fluid and Solid Mechanics'' (J.M.Burgerscentrum) for
financial support; and the institute of ``Mesoscopic Transport Phenomena''
(Technical University Eindhoven) for their hospitality. Further thanks go to the
Bayerische Forschungsstiftung and KONWIHR project waLBerla-EXA for financial
support.

\bibliography{lit}

\end{document}